\documentclass[prl,twocolumn,superscriptaddress]{revtex4-2}

\usepackage[utf8]{inputenc}
\usepackage{amsmath,amssymb,amsfonts,bm}
\usepackage{braket}
\usepackage[colorlinks=true,linkcolor=blue,urlcolor=blue,citecolor=blue,anchorcolor=blue]{hyperref}
\usepackage[capitalise]{cleveref}
\usepackage{graphicx}
\usepackage{float}
\usepackage{xcolor}
\usepackage{orcidlink}
\usepackage{comment}

\graphicspath{{figs/}}

\Crefname{section}{Sec.}{Secs.}
\crefrangelabelformat{equation}{(#3#1#4--#5#2#6)}

\newcommand{\dszero}{\ensuremath{D_{s0}^*} }
\newcommand{\dsone}{\ensuremath{D_{s1}} }
\newcommand{\bszero}{\ensuremath{B_{s0}^*} }
\newcommand{\bsone}{\ensuremath{B_{s1}} }

\newcommand{\sect}[1]{{\it #1}.---}

\begin{document}
%================ Basic info ================%
\title{Threshold cusp effects to measure masses of radiatively decaying hadrons:\\ The $B_{s0}^*$ mass from the $\Upsilon\phi$ spectrum}

\newcommand{\ITP}{\affiliation{Institute of Theoretical Physics,
Chinese Academy of Sciences, Beijing 100190, China }}
\newcommand{\UCAS}{\affiliation{School of Physical Sciences, University of Chinese Academy of Sciences (UCAS), Beijing 100049, China}}

\newcommand{\peng}{\affiliation{Peng Huanwu Collaborative Center for Research and Education, Beihang University, Beijing 100191, China}}

\newcommand{\fzj}{\affiliation{Institute for Advanced Simulation, Forschungszentrum J\"ulich, D-52425 J\"ulich, Germany}}

\newcommand{\swu}{\affiliation{School of Physical Science and Technology, Southwest University, Chongqing 400715, China}}

\newcommand{\scnt}{\affiliation{Southern Center for Nuclear-Science Theory (SCNT), Institute of Modern Physics,\\ Chinese Academy of Sciences, Huizhou 516000, China}}

\newcommand{\hiskp}{
\affiliation{Helmholtz--Institut f\"ur Strahlen- und Kernphysik (Theorie) 
and Bethe Center for Theoretical Physics,\\ Universit\"at Bonn, D-53115 Bonn, Germany}}

 \author{Hai-Long Fu\orcidlink{0000-0002-1722-4145}}
 \hiskp \fzj

 \author{Xu Zhang\orcidlink{0000-0002-3687-248X}}\email[Corresponding author: ]{zhangxu@itp.ac.cn}
 \ITP
 
 \author{Feng-Kun Guo\orcidlink{0000-0002-2919-2064}}\email[Corresponding author: ]{fkguo@itp.ac.cn}
 \ITP \UCAS \scnt
 
 \author{Christoph Hanhart\orcidlink{0000-0002-3509-2473}} 
 \fzj 

 \author{Ulf-G.~Mei{\ss}ner\orcidlink{0000-0003-1254-442X}}
\hiskp\fzj\peng
 
 \author{Mao-Jun Yan\orcidlink{0000-0002-3539-6514}}
 \swu

\begin{abstract}

Hadrons that decay predominantly into final states containing photons are notoriously difficult to detect at hadron colliders. Prominent examples are the yet-unobserved $B_{s0}^*$ and $B_{s1}$, the bottom partners of the $D_{s0}^*(2317)$ and $D_{s1}(2460)$, which are expected to exhibit exotic properties deviating from the conventional quark-model predictions of $\bar b s$ mesons. We propose a general, model-independent method to overcome this problem: when the target hadron has an attractive $S$-wave interaction with a companion hadron of precisely known mass, the line shape of a suitable final state develops a cusp at the pair threshold, or a peak just below it if the attraction binds, so that subtracting the companion mass returns the target mass, up to the binding energy in the latter case. As a proof of concept, a leading-order particle-dimer calculation of the $D\bar D_s K$ three-body system reproduces the $X(4274)$ structure in the LHCb $J/\psi\phi$ distribution extracted from $B\to K J/\psi \phi$,  as a $D_{s0}^*\bar D_s$ threshold cusp driven by a nearby virtual-state pole, yielding $m_{D_{s0}^*}=(2322\pm6)$~MeV in agreement with its measured value and favoring $J^{PC}=0^{-+}$ for the $X(4274)$. Transferring the elastic three-body interaction to the bottom sector, we predict an analogous structure at the $B_{s0}^*\bar B_s$ threshold near $11.09$~GeV, making the $\Upsilon\phi$ invariant-mass distribution at the LHC a clean probe of the $B_{s0}^*$ mass.

\end{abstract}

\maketitle

\sect{Introduction}
Experiments at the Large Hadron Collider (LHC), in particular with the LHCb experiment, have substantially advanced hadron spectroscopy by uncovering a variety of hadrons containing heavy quarks that are difficult to access at lower-energy facilities. Notable examples include the doubly charmed baryons~\cite{LHCb:2017iph,LHCb:2026pxn}, the hidden-charm $P_c$ pentaquark states~\cite{LHCb:2015yax,LHCb:2019kea}, and the doubly charmed tetraquark $T_{cc}^+$~\cite{LHCb:2021vvq}. These discoveries share a common feature: they were made in final states that are relatively easy to detect, i.e., free of photons.
Photon detection at the LHC is challenging because of limited energy resolution and large backgrounds.

A method to measure the masses of hadrons whose decay products  include at least one photon would therefore be highly valuable. Consider the example of the $\dszero(2317)$ and $\dsone(2460)$, discovered in 2003 by the BaBar~\cite{BaBar:2003oey} and CLEO~\cite{CLEO:2003ggt} Collaborations, respectively. These discoveries sparked numerous theoretical investigations because their masses lie significantly below quark-model predictions, the $D^\ast_{s0}(2317)$ is about $160\ \text{MeV}$ below the predicted mass for the lowest-lying $c\bar s$ scalar meson~\cite{Godfrey:1985xj,Godfrey:2015dva} (see also Refs.~\cite{Lakhina:2006fy,Ortega:2016mms}). Various models have been proposed, describing them as primarily $c\bar{s}$ states~\cite{Dai:2003yg,Narison:2003td,Lee:2004gt,Wang:2006bs}, as chiral partners of the ground-state $D_s^{(*)}$ mesons in the chiral parity doublet model~\cite{Bardeen:2003kt,Nowak:2003ra}, mixtures of $c\bar{s}$ with tetraquark~\cite{Browder:2003fk} or meson-meson~\cite{vanBeveren:2003kd,Albaladejo:2018mhb,Yang:2021tvc} components, purely tetraquarks~\cite{Cheng:2003kg,Terasaki:2003qa,Maiani:2004vq,Bracco:2005kt,Wang:2006uba}, or as heavy-light meson molecules~\cite{Barnes:2003dj,Szczepaniak:2003vy,Kolomeitsev:2003ac,Hofmann:2003je,Chen:2004dy,Guo:2006fu,Guo:2006rp,Gamermann:2006nm,Faessler:2007gv,Flynn:2007ki,Albaladejo:2016hae,Fu:2021wde}, driven by the proximity of the $D^\ast_{s0}(2317)$ to the $DK$ threshold and of the $D_{s1}(2460)$ to the $D^*K$ threshold.

Heavy-quark flavor symmetry predicts the bottom partners $\bszero$ and $\bsone$ of these two $D_{sJ}$ mesons  model-independently to be about 5.71 and 5.76~GeV, respectively~\cite{Fu:2021wde}, in agreement with results from lattice quantum chromodynamics (QCD)~\cite{Lang:2015hza,Hudspith:2023loy}. These values are consistent with predictions from the chiral doublet model~\cite{Bardeen:2003kt} and for $BK$ and $B^*K$ molecules~\cite{Kolomeitsev:2003ac, Guo:2006fu,Guo:2006rp,Albaladejo:2015lob,Du:2017zvv,Fu:2021wde}, but substantially smaller than the Godfrey-Isgur quark model prediction (5831 and 5857~MeV, respectively)~\cite{Godfrey:2016nwn}. This sizable spread among the predicted masses makes it important to search for $\bszero$ and $\bsone$ and to determine their masses experimentally.

However, most decay channels of the $\bszero$ and $\bsone$ involve one or two photons in the final state. These include $\bszero\to B_s\pi^0$ (with $\pi^0\to\gamma\gamma$) and $\bszero\to B_s^*\gamma$ for the $\bszero$, and $\bsone\to B_s^*\pi^0$, $B_s\gamma$, $B_s^*\gamma$, and $\bszero\gamma$ for the $\bsone$, which, however, also has a two-pion decay channel, $\bsone\to B_s\pi^+\pi^-$ but with a partial width as small as $(3\pm1)$~keV~\cite{Tang:2023yls}. It is therefore challenging to find the $\bszero$ and $\bsone$ in the LHC environment.

In this Letter, we propose to employ threshold cusp effects to search for signals of such particles and to measure their masses indirectly. Specifically, we suggest measuring the $\Upsilon\phi$ invariant-mass distribution in LHC experiments and searching for structures in the energy region from 11.0 to 11.2~GeV, i.e., around the $\bszero\bar{B}_s^{(*)}$ and $\bsone\bar{B}_s^{(*)}$ thresholds.
In particular, the $\bszero\bar B_s$ threshold is the lowest among them (line shapes around the higher thresholds can be more complicated~\cite{Zhang:2024qkg}); if the $\bszero\bar B_s$ $S$-wave interaction is attractive, there will be either a peak below the threshold when the attraction binds, or a cusp exactly at the threshold~\cite{Dong:2020hxe}.
In the latter case the line shape is asymmetric, with a shape that depends on the strength of the attraction.
We take the $\bszero\bar B_s$ system as an example in this Letter.
If the structure is a threshold cusp, the $\bszero\bar B_s$ threshold can be read off directly from the measured line shape, and the $\bszero$ mass follows model-independently by subtracting the well-measured $\bar B_s$ mass~\footnote{Since the $\bszero\bar B_s$ mesons that rescatter into $\Upsilon\phi$ are produced promptly at the LHC, complications due to the double-triangle singularity in $B\to J/\psi \phi K$ as discussed in Ref.~\cite{Nakamura:2021bvs} are avoided.}; if instead the attraction binds, the peak lies below the threshold by the binding energy, which must then be supplied by theory. In either case, an attractive $S$-wave interaction guarantees a nontrivial structure near the $\bszero\bar B_s$ threshold.

In this paper we demonstrate that the $J/\psi\phi$ distribution of $B^{+}\to J/\psi\phi K^{+}$ measured by the CDF and LHCb Collaborations~\cite{CDF:2011pep, LHCb:2016nsl, LHCb:2021uow} provides a proof of concept: a peak corresponding to the $X(4274)$ was observed around the $\dszero\bar{D}_s$ threshold~\cite{Dong:2020hxe,Nakamura:2023swt}.
Subtracting the $D_s$ mass from the $X(4274)$ mass, $(4290\pm6)$~MeV~\cite{ParticleDataGroup:2026}, we obtain $(2322\pm6)$~MeV, in perfect agreement with the mass of the $\dszero$, $(2317.8 \pm 0.5)$~MeV~\cite{ParticleDataGroup:2024cfk}.
If this peak is indeed due to the $\dszero\bar{D}_s$ threshold cusp, which is prominent because of a nearby pole, then the quantum numbers of the $X(4274)$ are $J^{PC}=0^{-+}$ rather than $1^{++}$ as suggested by the LHCb Collaboration on the basis of model-dependent analyses~\cite{LHCb:2016nsl,LHCb:2021uow}. Indeed, the LHCb data~\cite{LHCb:2021uow} around the $X(4274)$ peak show a cusp structure (see Fig.~\ref{fig:fit_scheme2} below).
Furthermore, around the almost degenerate $D_{s0}^*\bar D_s^*$ and $D_{s1}(2460)\bar D_s$ thresholds, there is a clear dip in the data between the reported $X(4274)$ and $X(4500)$~\cite{LHCb:2021uow}. Such a dip may also be connected to a threshold cusp in coupled-channel dynamics~\cite{Dong:2020hxe,Zhang:2024qkg}.

To work out the example explicitly, we study the $\dszero\bar{D}_{s}$ and $\bszero\bar{B}_{s}$ systems at leading order (LO) in the particle-dimer formulation of the nonrelativistic effective field theory (NREFT) for low-energy three-body interactions~\cite{Bedaque:1998kg,Bedaque:1998km}, for a textbook discussion, 
see~\cite{Meissner:2022cbi}.
We explore the pole structure generated from the three-body interaction and the resulting behavior of the near-threshold amplitude. At this order, the inputs to the Faddeev equations are the subsystem two-body scattering lengths and an additional three-body contact term. 
We will show that the $J/\psi\phi$ distribution around the $\dszero\bar{D}_{s}$ threshold measured by LHCb~\cite{LHCb:2021uow} can be well reproduced by the NREFT calculation, and we will then make predictions for the $\bszero\bar{B}_{s}$ system.

\sect{Framework}We focus on the following system with isospin $I=0$ and positive $C$ parity:
\begin{equation}
    \ket{0^{-+}} = \frac{1}{\sqrt{2}}\left(\ket{D^{*}_{s0}\bar{D}_{s}} + \ket{\bar{D}^{*}_{s0}D_{s}}\right).
\end{equation}
All subsequent matrix elements are evaluated in this $C$-even combination.
Since the $D_{s0}^*(2317)$ appears as a pole in the isoscalar $DK$ scattering amplitude, the system can be treated as a $D\bar D_s K$ three-body system (see Ref.~\cite{Wu:2025fzx} for a recent prediction of an exotic $0^{--}$ state in this system) with the following channel decomposition:
\begin{align}
\ket{\dszero\bar{D}_{s}}&=\frac{1}{\sqrt{2}}(\ket{D^{0}K^{+}}\ket{\bar{D}_{s}}+\ket{D^{+}K^{0}}\ket{\bar{D}_{s}}),\nonumber\\
\ket{\sigma_{\bar{D}_{s}K}D}&=\frac{1}{\sqrt{2}}(\ket{\bar{D}_{s}K^{+}}\ket{D^{0}}+\ket{\bar{D}_{s}K^{0}}\ket{D^{+}}),
\nonumber\\
\ket{\sigma_{\bar{D}_{s}D}K}&=\frac{1}{\sqrt{2}}(\ket{\bar{D}_{s}D^{0}}\ket{K^{+}}+\ket{\bar{D}_{s}D^{+}}\ket{K^{0}}).
\end{align}

In the NREFT particle-dimer framework the Lagrangian is $\mathcal{L}=\mathcal{L}_{1}+\mathcal{L}_{2}+\mathcal{L}_{3}$, with $\mathcal{L}_{1}$ the standard one-body kinetic terms for the pseudoscalar fields $\psi_a$ ($a=K,\bar{D}_{s},D$). The two-body sector is described by~\cite{Bedaque:1998kg,Bedaque:1998km}
\begin{equation}
\mathcal{L}_{2}=\sigma_{i}^{\dagger}\left(-i\partial_{0}+\Delta_{i}\right)\sigma_{i}-g_{i}\left(\sigma^{\dagger}_{i}\psi_{a}\psi_{b}+\text{h.c.}\right),
\end{equation}
where $\sigma_i$ is a dimer field with bare mass $\Delta_i$ and coupling $g_i$ for each $S$-wave pair $\psi_a\psi_b$ ($i=1,2,3$ for $DK$, $\bar D_s K$, and $\bar D_s D$, with the $DK$ dimer projected onto the $I=0$ singlet); the three particle--dimer channels are $(\sigma_{DK},\bar D_s)$, $(\sigma_{\bar D_sK},D)$, and $(\sigma_{\bar D_sD},K)$. At LO only the ratio $g_i^2/\Delta_i$ is physical, fixed by the two-body scattering length $a_i$. 

The dressed dimer propagator following from $\mathcal{L}_{2}$ is
\begin{equation}
\label{eq:dimeq}
    iG_{i}(E,\vec{p}) = \frac{-i\,{2\pi}/\left(\mu_{i}g^{2}_{i}\right)}{-1/a_{i} + \left[2\mu_{i}\left(\frac{\vec{p}^{2}}{2M_{i}} - E - i\epsilon\right)\right]^{1/2}},
\end{equation}
where $\mu_{i}$ is the reduced mass of the two particles coupled to the dimer $\sigma_i$, $M_i$ is the sum of their masses, and the scattering length $a_{i}$ is related to the two-body pole position at $-B_i$ relative to the corresponding threshold by $B_{i} = {1}/{(2\mu_{i}a_{i}^2)}$.
The denominator vanishes at the two-body pole $E_{\rm 2b}=-B_i$, on the physical (unphysical) sheet for $a_i>0$ ($a_i<0$); the wave-function renormalization $Z_i^{-1}=\mu_i^2 g_i^2 a_i/(2\pi)$ and the analytic continuation for virtual two-body states ($a_i<0$) are detailed in the Supplemental Material.

Finally, a three-body contact term $\mathcal{L}_3$ is needed to absorb the cutoff dependence in solving the three-body scattering equation with particle-exchange potentials~\cite{Bedaque:1998km,Bedaque:1998kg}.
For a recent discussion of charge-parity-constrained three-body forces in related hadronic systems, including $D\bar D_s K$, see Ref.~\cite{Pan:2025vyg}.

\begin{figure}[t]
\centering
\includegraphics[width=0.48\textwidth]{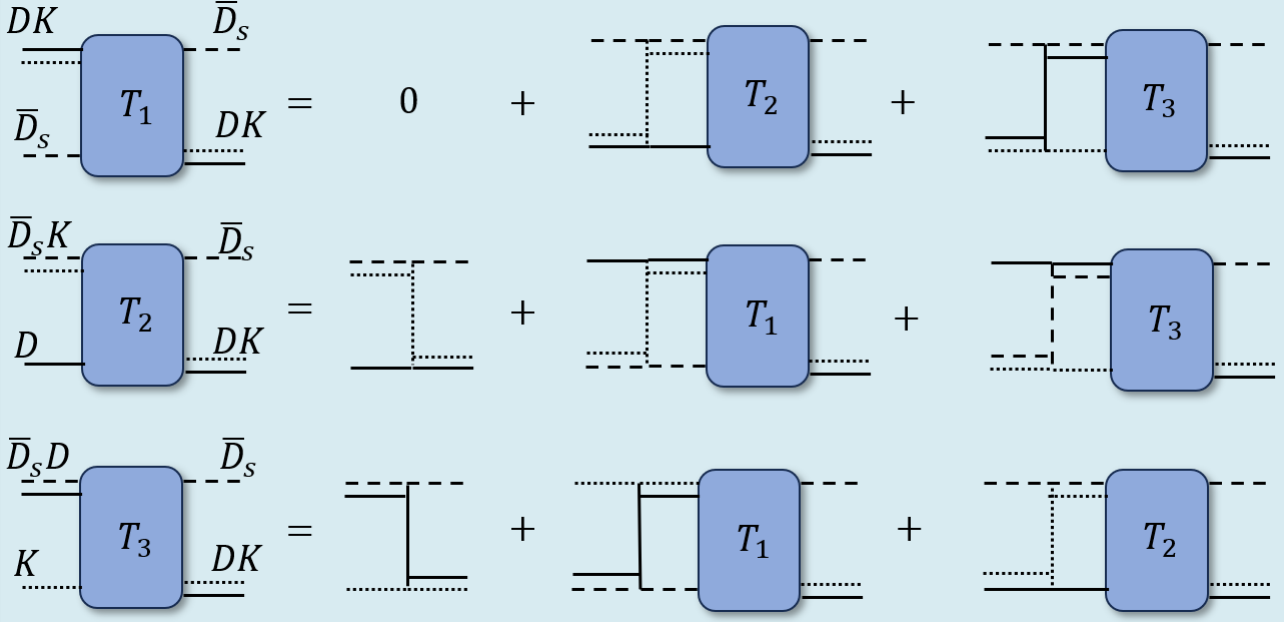}
\caption{Particle--dimer scattering equation without three-body force; single lines represent the spectator particle, while double lines depict the three dimers.}
\label{fig:LSE}
\end{figure}

The scattering equation for the positive $C$-parity $S$-wave $DK\bar D_s$ system in the dimer formalism is shown schematically in Fig.~\ref{fig:LSE}. 
In the absence of a three-body contact term it can be written as
\begin{align}
\label{eq:LSE}
\begin{pmatrix}
iT_{1}\\
iT_{2}\\
iT_{3}\\
\end{pmatrix}
=&
\begin{pmatrix}
0\\
iV_{21}\\
iV_{31}
\end{pmatrix} 
+\int^{\Lambda}_{0}\frac{k^{2}dk}{2\pi^{2}} \begin{pmatrix}
0&iV_{12}& iV_{13}\\
iV_{21}& 0& iV_{23}\\
iV_{31}& iV_{32}& 0
\end{pmatrix}
\nonumber \\
&\times\begin{pmatrix}
Z^{-1}_{1}iG_{1}&0&0\\
0& Z^{-1}_{2}iG_{2}& 0\\
0& 0& Z^{-1}_{3}iG_{3}
\end{pmatrix}
\begin{pmatrix}
iT_{1}\\
iT_{2}\\
iT_{3}
\end{pmatrix},
\end{align}
where $T_{i}$ is the particle-dimer amplitude with the $i$-th particle ($\bar{D}_{s}$, $D$, $K$) as the spectator, $T_{1}$ being the elastic $\dszero\bar{D}_{s}$ amplitude, and $\Lambda$ is a sharp cutoff.
The partial-wave projection of the potentials $V_{ij}$ generates a logarithmic branch cut that complicates the analytic structure of the scattering equation. We solve it by discretization and matrix inversion in momentum space~\cite{Haftel:1970zz}, using the contour-deformation method to handle this singularity, as discussed in Refs.~\cite{Aaron:1966zz,Hetherington:1965zza,Cahill:1971ddy,Glockle:1978zz,BookThree} (see also Ref.~\cite{Hu:2026f1}).
For the explicit form of the one-particle-exchange potential $V_{ij}$ and the technical details of the contour deformation, we refer to the Supplemental Material. 

\sect{Two-body scattering lengths} We work in the isospin limit with isospin-averaged masses $m_{D}=1867.23$~MeV and $m_{K}=495.66$~MeV. The two-body dynamics enters as input to the three-body calculation. At LO, the inputs are the scattering lengths $a_{i}$ in the two-body channels.
These scattering lengths have not been measured experimentally; however, reasonable estimates suffice as the main purpose of this Letter is not to make precise predictions of possible three-body poles, but rather to illustrate the use of a threshold cusp to measure the mass of a particle that is otherwise difficult to determine.

At LO of NREFT the scattering lengths in the two-body channels are tied to the molecular pole positions. Taking the masses of $\dszero$ $2317.8$~MeV~\cite{ParticleDataGroup:2024cfk} and its bottom partner $5722$~MeV predicted in Ref.~\cite{Fu:2021wde} as input, we can determine the scattering lengths $a_{DK}$ and $a_{BK}$.
The $\bar{D}_{s}K$ and $B_{s} K$ $S$-wave scattering lengths were given in a coupled-channel unitarized chiral perturbation theory (UChPT) framework~\cite{Liu:2012zya,Fu:2021wde}, in which there are $\bar D\pi$, $\bar D\eta$ ($B\pi$ and $B\eta$ in the bottom sector) channels with lower thresholds. Since we do not consider such inelastic channels here, the scattering lengths are computed for the single-channel $\bar{D}_{s}K$ and $B_{s} K$ scattering employing the same low-energy constants in UChPT as in Refs.~\cite{Liu:2012zya,Fu:2021wde}.

\begin{table}[t!]
    \caption{Scattering lengths used in this work. Here $H$ denotes $D$ ($\bar B$) in the charm (bottom) sector, and $a_D$ ($a_B$) the corresponding scattering lengths. For the $\bar H_s H$ channel the two entries are extracted from the cutoff-dependent contact interaction of Ref.~\cite{Ji:2022uie} (see also Ref.~\cite{Yan:2021tcp}) at $\Lambda_2=0.5$ and $1.0$~GeV, respectively, which we treat in parallel.}
    \begin{ruledtabular}
    \begin{tabular}{cccc}
    Channel & $HK$ & $\bar{H}_{s}K$ & $\bar{H}_{s}H$ \tabularnewline
    \hline 
    $a_{D}$ (fm) &   $1.05$ & $-2.57$& $0.22,\,-0.45$
    \tabularnewline
    $a_{B}$ (fm) &  $0.90$ & $-5.73$& $0.44,\,2.34$
    \end{tabular}
\end{ruledtabular}
    \label{Tab:InputB}
\end{table}

The $\bar{D}_{s}D$ interaction is not as well known as that between a charmed and a light pseudoscalar meson. Nevertheless, there exist reasonable predictions in the literature.
The results of Refs.~\cite{Ji:2022uie,Yan:2021tcp} indicate that the interaction between $\bar{D}_{s}$ and $D$ is not strong enough to form a bound state but can produce a virtual state.
In Ref.~\cite{Ji:2022uie} the $\bar{D}_{s}D$ interaction was described by a Lippmann-Schwinger equation with a cutoff-dependent contact potential $C_{1a}$, whose value was determined to be $C_{1a}=0.36^{+0.25}_{-0.26}$~fm$^{2}$ for $\Lambda_{2}=0.5$~GeV and $C_{1a}=-0.31^{+0.03}_{-0.05}$~fm$^{2}$ for $\Lambda_{2}=1.0$~GeV, where $\Lambda_2$ is the cutoff of the Gaussian form factor used for the two-body interaction in Ref.~\cite{Ji:2022uie}, distinct from the sharp three-body momentum cutoff $\Lambda$ in Eq.~\eqref{eq:LSE}. 
Using the same formalism, we extract the scattering lengths $a_{\bar{D}_{s}D}=0.22^{+0.10}_{-0.15}$~fm for $\Lambda_{2}=0.5$~GeV and $a_{\bar{D}_{s}D}=-0.45^{+0.07}_{-0.16}$~fm for $\Lambda_{2}=1.0$~GeV, respectively. The two extractions reflect the strong regulator dependence of this poorly constrained interaction: at $\Lambda_{2}=1.0$~GeV the contact is attractive ($C_{1a}<0$), giving a $\bar{D}_{s}D$ virtual-state pole below threshold, whereas at $\Lambda_{2}=0.5$~GeV there is no near-threshold pole. We will use both values and show that the threshold cusp, on which the proposed mass measurement relies, remains at the $\dszero\bar{D}_{s}$ threshold in both cases.
There is no heavy-quark flavor symmetry in the doubly-heavy system at the observable level~\cite{Baru:2018qkb}; however, the potential from exchanging light mesons is expected to follow the symmetry~\cite{Guo:2013sya}. We therefore use the same $C_{1a}$ value to estimate the scattering length $a_{\bar{B}_{s}B}$ in the bottom sector.
We obtain $a_{\bar{B}_{s}B}=0.44^{+0.14}_{-0.27}$~fm and $a_{\bar{B}_{s}B}=2.34^{+1.42}_{-0.77}$~fm at $\Lambda_{2}=0.5$~GeV and $\Lambda_{2}=1.0$~GeV, respectively. 
All the scattering lengths used in the following calculations are listed in Table~\ref{Tab:InputB}.

\sect{Results for $D\bar D_s K$}
The Faddeev equation can be solved to search for poles in the three-body scattering amplitude. Only the two Riemann sheets connected by the $\dszero\bar{D}_{s}$ (i.e., $\sigma_{1}\bar{D}_{s}$) unitary cut are relevant: the physical sheet and the adjacent unphysical one. To reach the latter, we need the discontinuity across the cut and get the second-sheet amplitude $T_{1}^{\rm II}=T_{1}^{\rm I}/(1+2i\rho_{1}\,T_{1}^{\rm I})$, where $\rho_{1}=\mu_{\rm bare} q_{\rm on}/(2\pi)$ is the two-body phase-space factor of the $\dszero\bar{D}_{s}$ pair, with $\mu_{\rm bare}=(m_D+m_K)m_{D_s}/(m_D+m_K+m_{D_s})$ the reduced mass appropriate to the dimer kinetic term and $q_{\rm on}$ the on-shell relative momentum of the $\dszero\bar{D}_{s}$ pair. A bound or virtual-state pole is located on the physical or unphysical Riemann sheet, respectively.

The Faddeev equation~\eqref{eq:LSE} is regularized with the sharp cutoff $\Lambda$.
We first set the three-body contact term to zero, taking $a_{\bar{D}_{s}D}=-0.45$~fm for illustration. Varying $\Lambda\in[0.2,1.0]$~GeV, we find a pole for the $J^{PC}=0^{-+}$ $D\bar D_s K$ system whose position depends strongly on the cutoff as expected. The pole is a bound state for $\Lambda\gtrsim \Lambda_{\star}\approx455\ \mathrm{MeV}$ and a virtual state for smaller values of $\Lambda$. We next include the three-body contact term, which absorbs the cutoff dependence to make all physical observables cutoff independent, and fit it to the LHCb $J/\psi\phi$ distribution.

\begin{figure}[tb]
    \centering
\includegraphics[width=\linewidth]{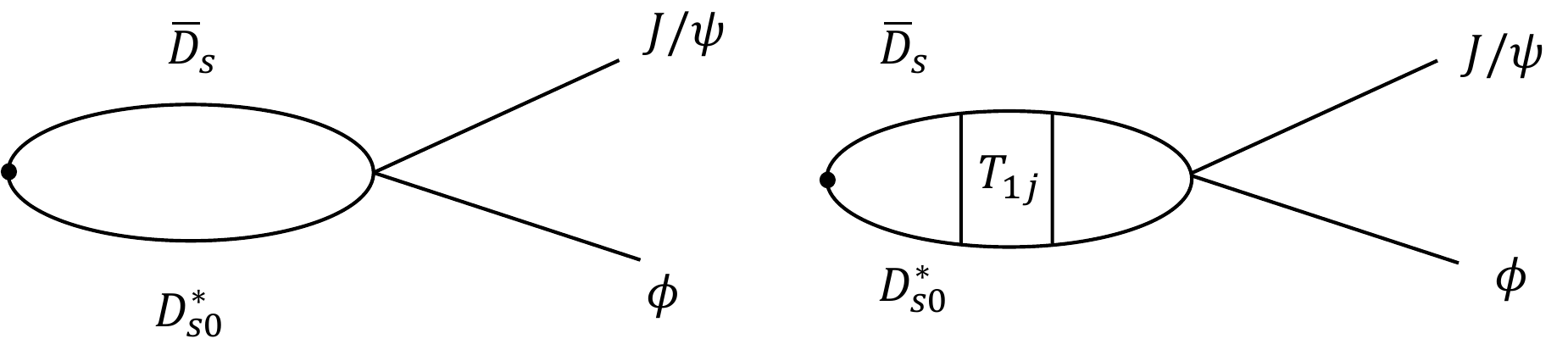}
    \caption{Production process of the $J/\psi\phi$ pair. Here $j=1,2,3$ labels the coupled channels $(D_{s0}^*,\bar D_s)$, $(\sigma_{\bar D_sK},D)$, $(\sigma_{\bar D_sD},K)$.}
    \label{fig:production}
\end{figure}

\sect{Fitting LHCb data with a three-body formalism}
To describe the $J/\psi\phi$ invariant-mass distribution measured by LHCb, we construct the production amplitude through the $\dszero\bar{D}_{s}$ intermediate state. We write
\begin{equation}
    iF=iF_1+iF_2\approx iF_{2},
\end{equation}
where $F_{1}$ is the direct $J/\psi\phi$ production amplitude, which is regular and thus negligible near the $\dszero\bar{D}_{s}$ threshold, so that $F\approx F_{2}$, which describes $J/\psi\phi$ production via the $\dszero\bar{D}_{s}$ intermediate state and the other two coupled channels; see Fig.~\ref{fig:production}.

These diagrams can be evaluated once the scattering matrix has been solved. The full form of $F_{2}$ at the energy $E\equiv\sqrt{s}-m_{D}-m_{D_{s}}-m_{K}$ reads
\begin{align}\label{eq:F2}
    F_{2}(E)&=U_{1}iV_{10}\int^{\tilde{\Lambda}}\frac{d^3k}{(2\pi)^3}iG_{1}\!\left(E-\frac{k^2}{2m_{D_{s}}},k\right)\nonumber \\ \nonumber
    &+U_{1}\sum_{j=1}^{3} iV_{j0}\int^{\tilde\Lambda}\frac{d^3k}{(2\pi)^3}\int^{\tilde\Lambda}\frac{d^3l}{(2\pi)^3}\\
    &\times iG_{1}\!\left(E-\frac{k^2}{2m_{D_{s}}},k\right)iT_{1j}(E,k,l)iG_{j}\!\left(E-\frac{l^2}{2m_{j}},l\right) \nonumber \\
    &\equiv U_1V_{10}f_{1}+U_1\sum_{j=1}^{3}V_{j0}f_{2}^{(j)},
\end{align}
where $m_j=(m_{D_{s}},m_{D},m_{K})$ are the spectator masses in $(G_{1},G_{2},G_{3})$, respectively.
Here we use the index $0$ for the $J/\psi\phi$ channel and $V_{j0}$ is the vertex from the dimer-spectator channel of $D\bar D_s K$ system to the $J/\psi\phi$ final state. $U_{1}$ is the prompt production amplitude for the near-threshold channel $\dszero\bar{D}_{s}$; subleading production through the other $D\bar D_s K$ channels is smooth in the near-$D_{s0}^*\bar D_s$-threshold window and is absorbed into the coherent background $b_0$ introduced below. 
The $G_{j}$ as given in Eq.~\eqref{eq:dimeq} denote the bubble diagrams shown in Fig.~\ref{fig:production}.  To regularize the integrations we use a sharp cutoff $\tilde \Lambda$, which for simplicity we set equal to the three-body cutoff $\Lambda$ in Eq.~\eqref{eq:LSE}.

The three-body contact term  $V^{\textrm{three}}_{ij}$ is introduced as
\begin{align}
iV^{\textrm{three}}_{ij}=i\mu_{i}g_i\,c\,g_j\mu_{j},
\label{eq:3bodycontact}
\end{align}
where $\mu_{i}$ and $\mu_{j}$ are the reduced masses in the incoming and outgoing $D\bar D_s K$ dimer-spectator channels $i$ and $j$.

Next we demonstrate that the LHCb data can be described by an amplitude built on the $D\bar D_s K$ three-body dynamics, and hence featuring a $D_{s0}^*\bar D_s$ threshold cusp. We take the short-range transition vertices $V_{j0}$ to be channel independent, which are then absorbed, together with the prompt-production factor $U_1$, into an overall normalization. 
Consequently, one gets the line-shape formula
\begin{align}\label{eq:3-body fitting 2}
     \frac{d N}{d M_{J/\psi\phi}}= \mathcal{N}^2\biggl|f_1+ \sum_{j=1}^3 f_2^{(j)} +b_{0}  \biggr|^2 {\rm p.s.},
\end{align}
where $f_{1}$ and $f_{2}^{(j)}$ are the integrals defined in the last line of Eq.~\eqref{eq:F2}, and $b_0=b_1+ib_2$ is a complex background term in the $0^{-+}$ channel that interferes with the resonant piece $f_1+\sum_{j}f_2^{(j)}$. The $P$-wave centrifugal factor (the $J/\psi\phi$ pair is in the $^3P_0$ partial wave required by $J^{PC}=0^{-+}$) is already included in the phase-space factor ${\rm p.s.}= \vert \vec{q}_{J/\psi\phi}\vert^3/(8\pi M_{J/\psi \phi})$ with $\vec{q}_{J/\psi\phi}$ the $J/\psi$ momentum in the $J/\psi\phi$ center-of-mass frame and $M_{J/\psi\phi}$ the $J/\psi\phi$ invariant mass. The full parameter set comprises the overall normalization $\mathcal{N}$, the background parameters $b_1$ and $b_2$, and the complex three-body coupling $c$.

We find that the LHCb data can be well described with Eq.~\eqref{eq:3-body fitting 2} by fixing $b_1=0$, as shown in Fig.~\ref{fig:fit_scheme2} for $a_{\bar{D}_{s}D}=0.22$~fm [(a), $\chi^2/\mathrm{d.o.f.}=2.3$] and $-0.45$~fm [(b), $\chi^2/\mathrm{d.o.f.}=3.0$], where $\Lambda=\tilde\Lambda=1$~GeV is taken; the corresponding amplitude has a virtual-state pole located at $(4266.2 - i\,1.7)$~MeV for $a_{\bar{D}_{s}D}=-0.45$~fm, whereas for $a_{\bar{D}_{s}D}=0.22$~fm no pole is found in the near-threshold region free of the one-particle-exchange left-hand cut (which sets in $26$~MeV below the $\dszero\bar{D}_{s}$ threshold; see the Supplemental Material), and the structure in the data is described by the threshold cusp alone.
Surprisingly, our fit agrees very well with the best-fit histogram reported in the original LHCb analysis~\cite{LHCb:2021uow}; for the details of fitting see Fig.~\ref{fig:fit_scheme2}.

\begin{figure}[tb]
    \centering
    \includegraphics[width=\linewidth]{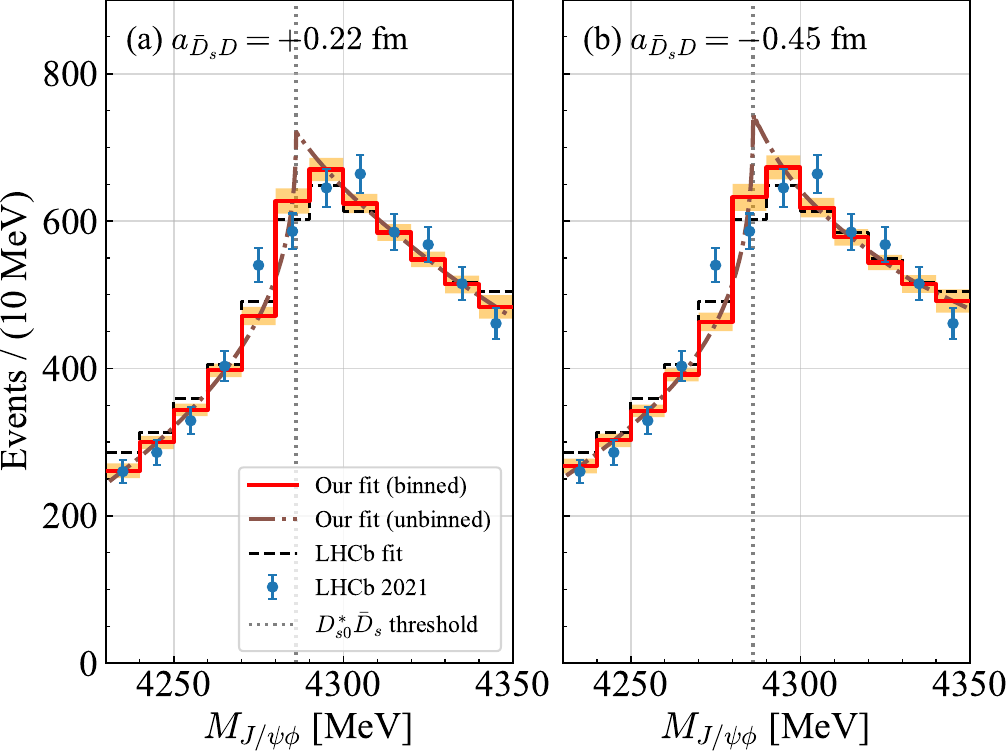}
    \caption{Fit to the LHCb data of the $J/\psi\phi$ invariant mass distribution in $B\to K J/\psi\phi$~\cite{LHCb:2021uow} with $\Lambda=\tilde\Lambda=1$~GeV, for (a) $a_{\bar{D}_{s}D}=0.22$~fm [$\chi^2/\mathrm{d.o.f.}=2.3$, $\mathcal{N}=0.0068\pm0.0005$ (arbitrary units), $b_0=(72\pm18)\,i$~MeV, $c=[(3.4\pm0.2)-i\,(3.9\pm0.1)]$~GeV$^{-3}$] and (b) $a_{\bar{D}_{s}D}=-0.45$~fm [$\chi^2/\mathrm{d.o.f.}=3.0$, $\mathcal{N}=0.041\pm0.004$, $b_0=(77\pm6)\,i$~MeV, $c=[(17.5\pm0.6)-i\,(11.0\pm0.2)]$~GeV$^{-3}$]. The red solid line is our bin-averaged fit with its error band and the brown dash-dotted line is the same fit before bin averaging; the black dashed line is the LHCb best-fit histogram~\cite{LHCb:2021uow}.}
    \label{fig:fit_scheme2}
\end{figure}

\sect{Results for $B\bar B_s \bar K$}With an analogous interaction, it is natural to expect the $0^{-+}$ $\bszero\bar{B}_{s}$ system to host a pole near threshold as well. Since we propose to search for the $\bszero$ indirectly through the $\bszero\bar{B}_{s}$ subsystem in the $\Upsilon\phi$ final state, we now estimate where this pole is located. We do this by transferring the three-body contact term in Eq.~\eqref{eq:3bodycontact} from the $D\bar{D}_{s}K$ system to the bottom analog with the corresponding reduced masses taken to be the bottom ones. 
Because the inelastic channels below threshold differ substantially between the two sectors, only the real part of $c$ extracted from the corresponding $J/\psi\phi$-invariant-mass-distribution fit is used to estimate the pole position in the bottom system.
Because no $\Upsilon\phi$ data are yet available to fix the bottom three-body term and thereby reabsorb the cutoff dependence as in the charm sector, we instead vary the cutoff ($\Lambda_b$) in Eq.~\eqref{eq:LSE} over a representative range. 
For the range $\Lambda_b \in[0.75,1.50]$~GeV, there is always a pole near the $\bszero\bar{B}_{s}$ threshold as listed in Table~\ref{Tab:Pole of B with real c}: it is a virtual state at smaller $\Lambda_b$ and crosses onto the physical sheet to become a shallow bound state at $\Lambda_{b}\simeq 1.32$~GeV ($1.24$~GeV) for $\Lambda_{2}=0.5$~GeV (1.0~GeV); decreasing $\Lambda_b$ further eventually pushes the virtual pole into the left-hand cut~\cite{Dawid:2023jrj}.

\begin{table}[tb]
    \caption{$\bszero\bar{B}_{s}$ pole positions as $\Lambda_{b}$ varies between $0.75$ and $1.50$~GeV. $\Delta E_B$ is the $\bszero\bar{B}_{s}$ pole position relative to the $\bszero\bar{B}_{s}$ two-body threshold; ``b" and ``v" denote bound and virtual states. In each entry the left (right) value corresponds using $\Lambda_{2}=0.5$~GeV ($1.0$~GeV), i.e., $a_{\bar{B}_{s}B}=0.44$~fm ($2.34$~fm) combined with $\mathrm{Re}\,c=3.4$~GeV$^{-3}$ ($17.5$~GeV$^{-3}$) from the corresponding $D\bar D_s K$ fit; only the real part of $c$ is used.}
    \begin{ruledtabular}
    \begin{tabular}{ccc}
      & pole position & $\Delta E_B$(MeV)
      \tabularnewline
    \hline
    $\Lambda_{b}=0.75$ GeV & v, v & $-9.28,\,-6.17$
    \tabularnewline
    $\Lambda_{b}=1.00$ GeV &  v, v & $-3.50,\,-1.39$
    \tabularnewline
    $\Lambda_{b}=1.25$ GeV & v, b & $-0.19,\,-0.005$
    \tabularnewline
    $\Lambda_{b}=1.50$ GeV & b, b & $-0.83,\,-1.44$
    \end{tabular}
\end{ruledtabular}
    \label{Tab:Pole of B with real c}
\end{table}

\sect{Summary}In summary, we have proposed that near-threshold cusp effects can be used as a general, model-independent route to measuring the masses of hadrons which are otherwise difficult to be observed. The required ingredients are
(i) an attractive $S$-wave interaction between the target hadron and a well-measured companion, and (ii) a clean experimental channel into which the corresponding pair can rescatter. 
The LHCb $J/\psi\phi$ data from $B^+\to J/\psi\phi K^+$~\cite{LHCb:2016axx,LHCb:2021uow} furnish a proof of concept in the charm sector: a particle-dimer LO NREFT calculation
of the $D\bar D_s K$ three-body system reproduces the structure at $\sim 4286$~MeV as a $\dszero\bar{D}_s$ threshold cusp
using two choices of $\bar{D}_{s}D$ interaction: for $a_{\bar D_s D}=-0.45$~fm the cusp is enhanced by a near-threshold virtual-state pole at $4266.2 - i\,1.7$~MeV, while for $a_{\bar D_s D}=+0.22$~fm no pole remains in the region free of the left-hand cut and the cusp alone describes the data.
This interpretation implies that the $X(4274)$ quantum numbers should be $J^{PC}=0^{-+}$, which need to be verified with a thorough analysis of the data considering coupled-channels effects.

Transporting the real (elastic) part of the fitted three-body contact term to the
bottom sector, we predict the existence of a pole in the $B\bar{B}_s\bar{K}$ three-body system close to the $B_{s0}^*\bar B_s$ threshold for the bottom-sector cutoff $\Lambda_b$ in a large range. 
The pole is either a shallow bound state  or a virtual-state pole below threshold. 
In both cases a nontrivial structure in the $\Upsilon\phi$ invariant mass distribution should exist at around $11.09$~GeV and encode the $\bszero$ mass, which can be obtained by subtracting the $B_s$ mass out model-independently from the cusp position at the $\bszero\bar{B}_s$ threshold for a virtual state, or up to a theoretically estimated binding energy for a bound state.

The method extends to any hadron which
interacts attractively in the $S$ wave with a companion of precisely known mass.
A measurement of the $\Upsilon\phi$ invariant-mass spectrum in the $11.0$--$11.2$~GeV
region at the LHC would constitute a direct and model-independent probe of the
$\bszero$ mass, thus offering crucial information for understanding heavy-flavor exotic hadrons and heavy-meson--light-meson interactions.

\medskip
\begin{acknowledgments}
    This work is supported in part by the National Natural Science Foundation of China under Grants No. 12125507, No. 12361141819, and No. 12447101; 
    by  the National Key R\&D Program of China under Grant No. 2023YFA1606703;
    by the Chinese Academy of Sciences (CAS) under Grants No.~YSBR-101, and 
    by Deutsche Forschungsgemeinschaft (DFG) under Grant No. 525056915
    under Germany's Excellence Strategy -- EXC 3107 -- Project-ID~533766364. 
    C.H. and U.-G.M. also acknowledge the support from the CAS President's International Fellowship Initiative (PIFI) under Grants No.~2025PD0087 and No.~2025PD0022, respectively.

\end{acknowledgments}

\bibliography{refs.bib}

%========================================================
%		Supplemental Material
%========================================================
\newpage

% %%%%%%%%%%%%%%%%%%%%%%%%%%%%%%%%%%%%%%%%%%%%%%%%%%%%%%%%%%%%%%%%%%%%%%%%%%%%%%%%%%%%%%%%%%%%%%%%%%%

\def\modified#1{\red{#1}}

\onecolumngrid
\section{Supplementary Materials}
\label{sup}

\setcounter{equation}{0}
\setcounter{figure}{0}
\setcounter{table}{0}
\renewcommand{\theequation}{S\thesection\arabic{equation}}
\renewcommand{\thefigure}{S\thesection\arabic{figure}}
\renewcommand{\thetable}{S\thesection\arabic{table}}

This supplemental material includes additional information on the details of the amplitude calculation and the contour-deformation method used to treat the
logarithmic singularities arising from the partial-wave projection of the potential.

\subsection*{S1. Scattering equation}
The two-body $S$-wave interactions are described by the dimer Lagrangian~\cite{Bedaque:1998kg,Bedaque:1998km}
\begin{equation}
    \mathcal{L}_{2} = \sigma_{i}^{\dagger} \left(-i\partial_{0}  + \Delta_{i}\right)\sigma_{i} - g_{i} \left(\sigma^{\dagger}_{i}\psi_{a}\psi_{b} + \text{h.c.}\right)
    \label{eq:L2},
\end{equation}
where $\sigma_{i}$ ($i=1,2,3$ for $DK$, $\bar{D}_{s}K$, and $\bar{D}_{s}D$) is the dimer field coupling to the corresponding meson pair and $g_i,\Delta_i$ are two-body parameters; for the $DK$ dimer the $I=0$ projection gives
\begin{equation}
    \mathcal{L}_{DK} = -\frac{g_{DK}}{\sqrt{2}} \sigma^{\dagger}_{DK}(\psi_{D^{0}}\psi_{K^{+}}+\psi_{D^{+}}\psi_{K^{0}}) + \text{h.c.}
    \label{eq:LDK}
\end{equation}

The scattering equation is depicted in Fig.~\ref{fig:LSE} of the main text, which for the $S$-wave is formulated as
\begin{align}
\label{eq:LSES}
\begin{pmatrix}
iT_{1}(p,q)\\
iT_{2}(p,q)\\
iT_{3}(p,q)\\
\end{pmatrix}
=
\begin{pmatrix}
0\\
iV_{21}(p,q)\\
iV_{31}(p,q)
\end{pmatrix} 
+\int^{\Lambda}_{0}\frac{k^{2}dk}{2\pi^{2}} \begin{pmatrix}
0&iV_{12}(p,k)& iV_{13}(p,k)\\
iV_{21}(p,k)& 0& iV_{23}(p,k)\\
iV_{31}(p,k)& iV_{32}(p,k)& 0
\end{pmatrix}
\begin{pmatrix}
Z^{-1}_{1}iG_{1}(E-\frac{k^{2}}{2m_{D_{s}}},k)iT_{1}(k,q)\\
Z^{-1}_{2}iG_{2}(E-\frac{k^{2}}{2m_{D}},k)iT_{2}(k,q)\\
Z^{-1}_{3}iG_{3}(E-\frac{k^{2}}{2m_{K}},k)iT_{3}(k,q)
\end{pmatrix},
\end{align}
where $p$ and $q$ denote the outgoing and incoming spectator momenta, respectively.
The dressed dimer propagator following from $\mathcal{L}_{2}$ reads
\begin{equation}
\label{eq:Gi}
    iG_{i}(E,\vec{p}) = \frac{-i\,{2\pi}/\left(\mu_{i}g^{2}_{i}\right)}{-1/a_{i} + \left[2\mu_{i}\left(\frac{\vec{p}^{2}}{2M_{i}} - E - i\epsilon\right)\right]^{1/2}},
\end{equation}
where $\mu_{i}$ ($M_{i}$) is the reduced mass (the mass sum) of the two particles forming the dimer $\sigma_{i}$, and $a_{i}$ is the corresponding two-body scattering length, related to the two-body binding energy by $B_{i}=1/(2\mu_{i}a_{i}^{2})$; the wave-function renormalization is $Z_{i}^{-1}=\mu_{i}^{2}g_{i}^{2}a_{i}/(2\pi)$. The one-particle-exchange potentials read
\begin{align}
    &V_{21}(\vec{p},\vec{q},E)=\frac{-\sqrt{Z_{2}}\sqrt{Z_{1}}g_{2}g_{1}}{E-\frac{p^{2}}{2m_{D}}-\frac{q^{2}}{2m_{\bar{D}_{s}}}-\frac{(\vec{q}+\vec{p})^{2}}{2m_{K}}+i\epsilon} , \nonumber\\
    &V_{31}(\vec{p},\vec{q},E)=\frac{-\sqrt{Z_{3}}\sqrt{Z_{1}}g_{3}g_{1}}{E-\frac{p^{2}}{2m_{K}}-\frac{q^{2}}{2m_{\bar{D}_{s}}}-\frac{(\vec{q}+\vec{p})^{2}}{2m_{D}}+i\epsilon},  \nonumber\\
    &V_{23}(\vec{p},\vec{q},E)=\frac{-\sqrt{Z_{2}}\sqrt{Z_{3}}g_{2}g_{3}}{E-\frac{p^{2}}{2m_{D}}-\frac{q^{2}}{2m_{K}}-\frac{(\vec{q}+\vec{p})^{2}}{2m_{\bar{D}_{s}}}+i\epsilon},
\end{align}
and $V_{ji}(\vec{p},\vec{q})=V_{ij}(\vec{q},\vec{p})$. The $S$-wave projection of the interaction potential can be written as
\begin{equation}
\label{eq:partPoten}
V_{ij}(p,q,E)=\frac{1}{2}\int_{-1}^{+1}d(\cos{\theta}) \frac{-\sqrt{Z_{i}}\sqrt{Z_{j}}g_{i}g_{j}}{E-\frac{p^{2}}{2m_{1}}-\frac{q^{2}}{2m_{2}}-\frac{p^2+q^2+2pq\cos{\theta}}{2m_{3}}+i\epsilon},
\end{equation}
where $\theta$ is the angle between the incoming momentum $\vec{q}$ and the outgoing momentum $\vec{p}$. Here $m_1,m_2,m_3$ denote the outgoing-spectator, incoming-spectator, and exchange-particle masses, respectively; for $V_{12}$ this is $(m_1,m_2,m_3)=(m_{\bar D_s},m_D,m_K)$, and for $V_{13}$, $(m_1,m_2,m_3)=(m_{\bar D_s},m_K,m_D)$.
The final result used in Eq.~\eqref{eq:LSES} is
\begin{equation}
V_{ij}(p,q,E)=\sqrt{Z_{i}}\sqrt{Z_{j}}g_{i}g_{j}\frac{m_3}{2pq}\log\frac{-E+\frac{p^2}{2m_1}+\frac{q^2}{2m_2}+\frac{(p+q)^2}{2m_3}-i\epsilon}{-E+\frac{p^2}{2m_1}+\frac{q^2}{2m_2}+\frac{(p-q)^2}{2m_3}-i\epsilon}.
\end{equation}

At LO, the three-body contact interaction may be written as
\begin{equation}\label{eq:Vthree}
V^{\rm three}_{ij}(p,q,E)=\mu_i\,g_i\,c(\Lambda)\,g_j\,\mu_j,
\end{equation}
where $c(\Lambda)$ is the running three-body coupling fixed by a renormalization condition~\cite{Bedaque:1998kg,Bedaque:1998km}. When the contact term is included, the kernel potential $V_{ij}$ in Eq.~\eqref{eq:LSES} is replaced by $V_{ij}+V^{\rm three}_{ij}$ for all channel pairs, including the diagonal ones. 

\subsection*{S2. Numerical method}

\begin{figure}[tbh]
    \centering
    \includegraphics[width=0.495\textwidth]{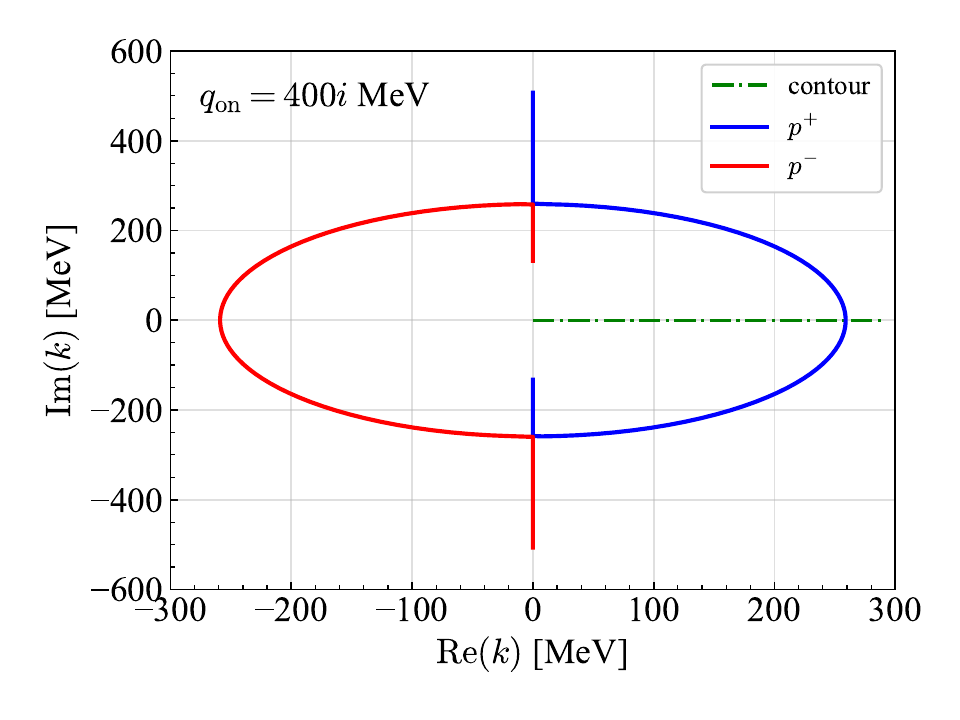}\hfill
    \includegraphics[width=0.495\textwidth]{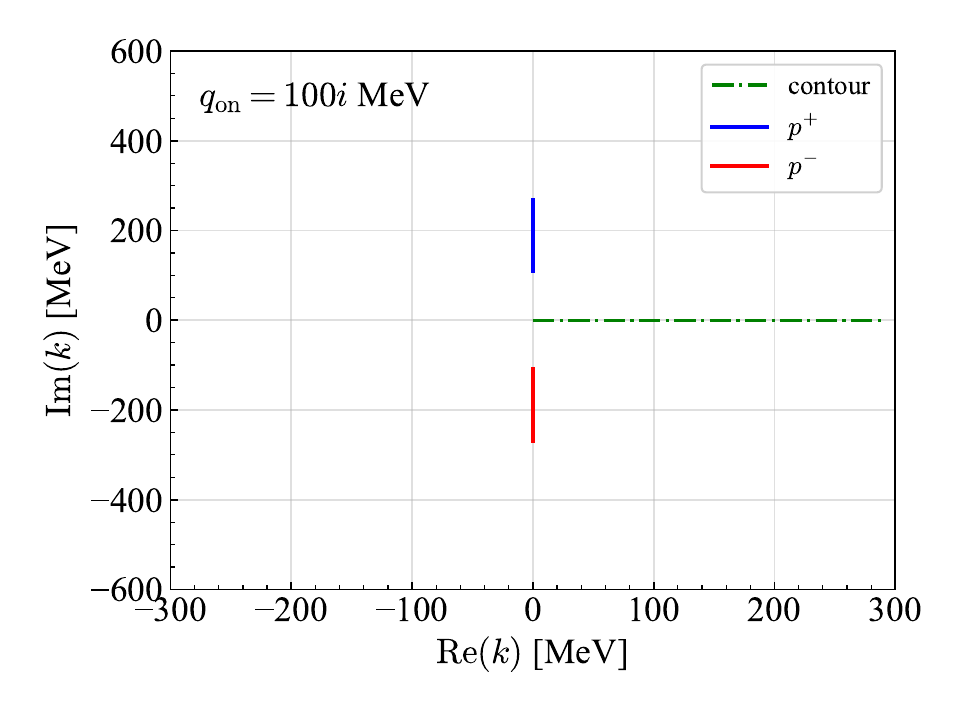}\\
    \includegraphics[width=0.495\textwidth]{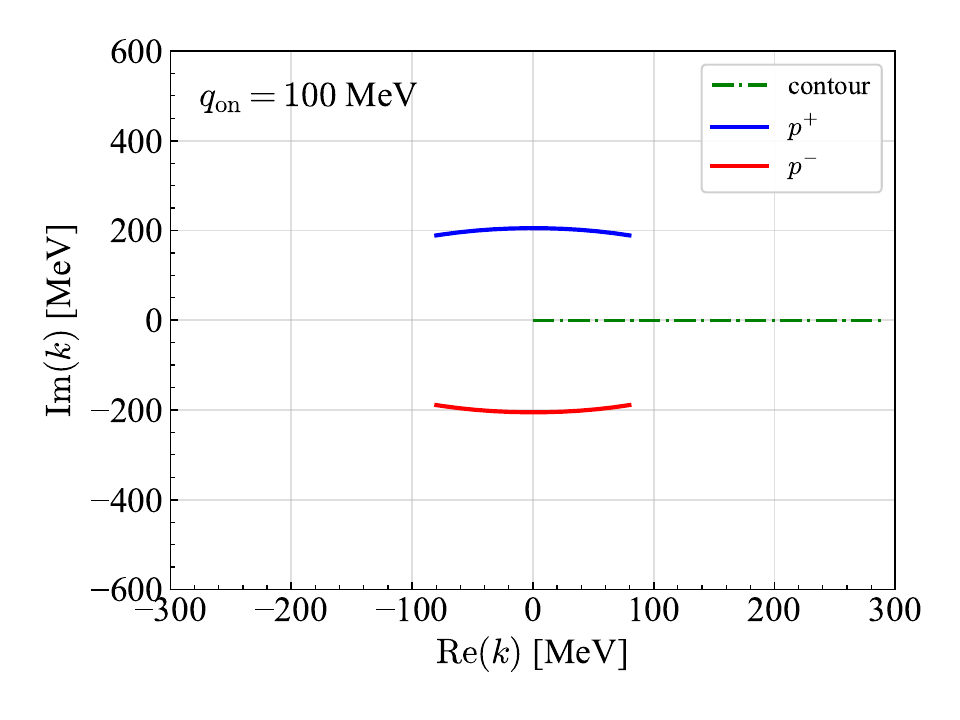}
    \caption{The $k$-plane cut of half on-shell $V_{12}$, evolving with $q_{\rm on}$ near threshold. We take $q_{\rm on}=400i,\ 100i,\ \text{and}\ 100$~MeV to show the deformation of the one-particle-exchange cut. The unit of axis is MeV. Blue lines and red lines represent $p^{+}$ and $p^{-}$ respectively. Green dashed dot line is a naive integration contour.}
    \label{fig:onshell}
\end{figure}

The integral equation Eq.~\eqref{eq:LSES} is solved numerically by discretizing the integrals with Gauss--Legendre quadrature and inverting the resulting linear system~\cite{Haftel:1970zz}. We first choose a suitable integration contour and discretize the momentum integral, casting the scattering equation into matrix form (channel indices and the $Z_i^{-1}$ factor preceding $G_i$ are suppressed for clarity):
\begin{align}    T(p_{i},q_{\rm on},E)&=V(p_{i},q_{\rm on},E)+\sum_{j}\frac{k^2_{j}}{2\pi^2} V(p_{i},k_{j},E)G_j(k_{j},E)T(k_{j},q_{\rm on},E)\omega_{j}\\ &=V(p_{i},q_{\rm on},E)+\sum_{j}K(p_{i},k_{j},E)T(k_{j},q_{\rm on},E)\omega_{j},\label{eq:half-onshellLSE}
\end{align}
the energy $E$ is fixed and the momentum $q$ is set on-shell,
\begin{equation}\label{eq:onshellE}
    q_{\rm on}=\sqrt{2\mu_{\rm bare}(E+B_{DK})},
\end{equation}
with $\mu_{\rm bare}=(m_D+m_K)m_{D_s}/(m_D+m_K+m_{D_s})$ the reduced mass of the $\dszero\bar D_s$ pair (the $\dszero$ enters at its bare dimer mass $m_D+m_K$; we write $\mu_{\rm bare}$ to distinguish it from the two-body reduced mass $\mu_i$ of the dimer constituents that enters $G_i$ and $Z_i$) and $-B_{DK}$ the $\dszero$ pole position relative to the $DK$ threshold. We take $q_{\rm on}$ on the first Riemann sheet, $\mathrm{Im}\,q_{\rm on}>0$. The resulting cut structure of $V_{12}$ in the complex $k$-plane is shown in Fig.~\ref{fig:onshell}; the analogue for $V_{13}$ is qualitatively similar but the circular cut appears far below the threshold. The momenta $p_{i}$ and $k_{j}$ are the Gauss--Legendre nodes, with weights $\omega_{j}$, and may in principle be complex. 

Solving Eq.~\eqref{eq:half-onshellLSE} by matrix inversion gives the half-on-shell solution $T_{\rm sol}=T(p_{i},q_{\rm on},E)$. In matrix notation,
\begin{equation}
[T]=[V]+[K][T]   , 
\end{equation}
which is solved by
\begin{equation}
[T]=(1-[K])^{-1}[V]    .
\end{equation}
Subsequently, we iterate this solution in the LSE to determine the $T$-matrix at the fixed energy $E$:
\begin{equation}
T(q_{\rm on},q_{\rm on},E)=V(q_{\rm on},q_{\rm on},E)+\sum_{i}K(q_{\rm on},k_{i},E)T_{\rm sol}(k_{i},q_{\rm on},E)\omega_{i}.
\end{equation}
For the channel-1 $\dszero\bar D_s$ right-hand cut, the sheet-II amplitude is obtained by adding the discontinuity below using the Cutkosky cutting rule, 
\begin{align}    
    T(p,q,E)&=V(p,q,E)+\delta VGT+\int \frac{k^2 dk}{2\pi^2} V(p,k,E)G(k,E)T(k,q,E),
\end{align}
where the discontinuity generated by placing the intermediate $\dszero\bar{D}_{s}$ pair on shell is
\begin{align}
\delta VGT=-2i\rho_{1}\,V(p,q_{\rm on},E)\,T(q_{\rm on},q,E),\qquad \rho_{1}=\frac{\mu_{\rm bare}\,q_{\rm on}}{2\pi},
\label{eq:Disc}
\end{align}
with $\mu_{\rm bare}$ the $\dszero\bar{D}_{s}$ reduced mass [Eq.~\eqref{eq:onshellE}], so that $\rho_1$ is exactly the $\dszero\bar{D}_{s}$ phase-space factor that appears in the main text. Solving the discretized form of this equation gives the $T$-matrix on the second Riemann sheet; since the added term is exactly the unitarity discontinuity across the $\dszero\bar{D}_{s}$ cut, this construction reproduces the main-text second-sheet continuation $T_{1}^{\rm II}=T_{1}^{\rm I}/(1+2i\rho_{1}\,T_{1}^{\rm I})$.

The singularities of the propagator $G_{i}(E,p)$ [Eq.~\eqref{eq:Gi}] occur where its denominator vanishes,
\begin{equation}
-1/a_{i} + \left[2\mu_{i}\left(\frac{\vec{p}^{2}}{2M_{i}} - E - i\epsilon\right)\right]^{1/2}=0.
\end{equation}
Setting $p=0$, we obtain the two-body pole position in the $E$-plane,
\begin{equation}
E=-\frac{1}{2\mu_ia_i^2}.
\end{equation}
As $a_i$ becomes negative, this pole moves onto the unphysical Riemann sheet. These singularities determine the branch structure of the particle-dimer propagator; the physical channel threshold and the dimer-pole branch point should be distinguished. The different Riemann sheets of the amplitude are connected by a cut. 

\begin{figure}[tb]
    \centering
    \includegraphics[scale=0.6]{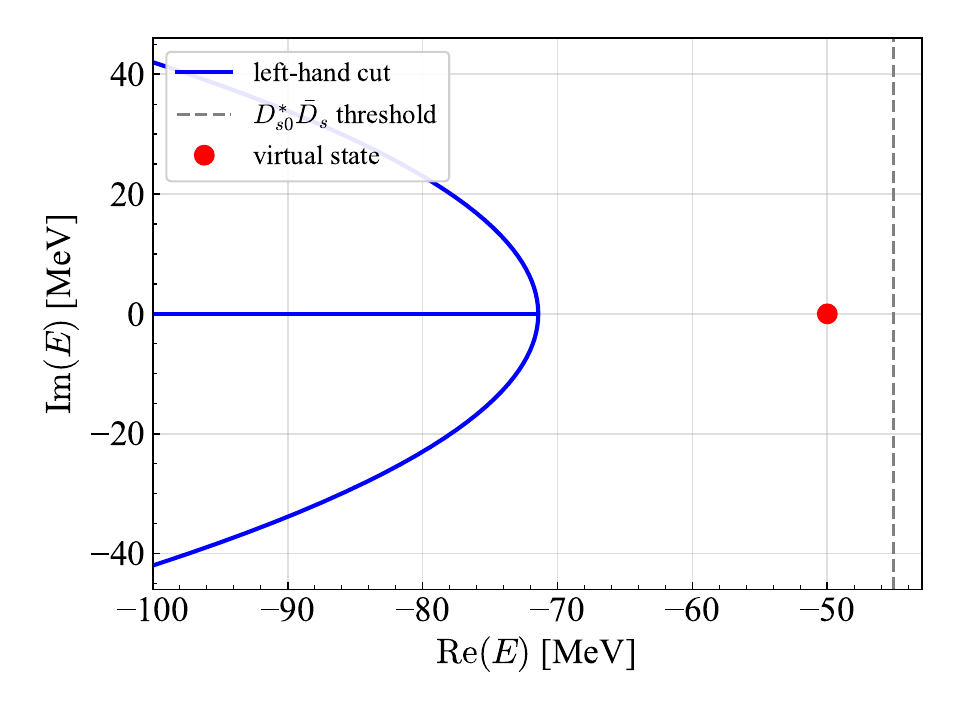}
    \caption{Analytic structure of $T$ in the $E$-plane. The red dot marks the possible position where a virtual pole may be found. If the three-body contact term $c$ is real, the virtual-state pole lies on the real axis of the second Riemann sheet, above the left-hand cut and below the $\dszero\bar D_s$ threshold. }
    \label{fig:cut str}
\end{figure}

Apart from the singularity in the propagator $G_{i}(E,p)$, there are singularities in the partial-wave projection of the potential $V_{ij}(p,q)$ in Eq.~\eqref{eq:partPoten}. The singularity occurs when the denominator of Eq.~\eqref{eq:partPoten} vanishes for any $\cos\theta\in [-1,+1]$. For a fixed three-body energy $E$ and an outgoing momentum $q$, the singularity of $V_{ij}$ is given as:
\begin{align}
    p^{\pm}(q,E,x)=\pm\frac{m_1m_3}{m_1+m_3}\sqrt{\frac{q^2 x^2}{m_{3}^2}+\frac{2(m_1+m_3)}{m_1m_3} \left(E-\frac{q^2}{2 m_{2}}-\frac{q^2}{2 m_{3}}\right)}-\frac{m_1m_3}{m_1+m_3}\frac{q x}{m_{3}}.
\end{align}
The cut of $V_{ij}(p,q)$ in the $q$-plane at fixed $p$ follows from the same expression with the mass ordering of $V_{ij}$ replaced by that of $V_{ji}$. In this work, we consider the on-shell scattering $\dszero\bar{D}_{s} \to \dszero\bar{D}_{s}$. The on-shell momentum can be obtained from Eq.~\eqref{eq:onshellE}. As an example, we consider the singularities in $V_{12}$, and the treatment of $V_{13}$ is analogous. Putting the outgoing momentum $q$ to be on-shell, the singularities in the complex $k$-plane corresponding to $E>-B_{DK}$ and $E<-B_{DK}$ are given in Fig.~\ref{fig:onshell}. From Fig.~\ref{fig:onshell}, one can see that for $E>-B_{DK}$ the singularities lie far from the real axis (bottom panel) and cause no difficulty for the integration. When $E_{\rm cir}<E<-B_{DK}$ the singularities lie on the imaginary axis as shown in the top-right panel of Fig.~\ref{fig:onshell}. As the value of $E$ decreases such that $E<E_{\rm cir}$, a circular cut will appear as shown in the top-left panel of Fig.~\ref{fig:onshell}. To obtain $E_{\rm cir}$, we set the cut endpoint to the origin, $p^{\pm}(q_{\rm on},E,1)=0$, which gives
\begin{equation}
   E_{\rm cir}=\frac{B_{DK} m_{D_s} (m_D+m_K) (m_a+m_b)}{m_a m_b m_{D_s}+(m_{D}+m_K)(m_a m_b -m_a m_{D_s} -m_b m_{D_s})}.
\end{equation}
Here $m_a$ and $m_b$ denote the two masses entering the circular-cut condition of $V_{ij}$ (distinct from the projection masses $m_{1,2,3}$ in Eq.~\eqref{eq:partPoten}): for $V_{12}$, $(m_a,m_b)=(m_K,m_{D_s})$; for $V_{13}$, $(m_a,m_b)=(m_D,m_{D_s})$.
For $V_{12}$ we get $E_{\rm cir}=-71.4$~MeV, while for $V_{13}$ we get $E_{\rm cir}=-418.9$~MeV so that the circular cut of $V_{13}$ does not affect the near-threshold region considered here. These values are obtained with the isospin-averaged masses $m_D=1867.23$~MeV, $m_K=495.66$~MeV, $m_{D_s}=1968.34$~MeV, $m_{D_{s0}^*}=2317.8$~MeV~\cite{ParticleDataGroup:2024cfk}, and $B_{DK}\approx 45.1$~MeV. The resulting analytic structure of $T$ in the complex $E$-plane---the $\dszero\bar{D}_{s}$ threshold, the left-hand cut setting in at $E_{\rm cir}$, and the second-sheet region in which a virtual-state pole may appear for a real three-body contact term $c$---is shown in Fig.~\ref{fig:cut str}.

\begin{figure}[]
    \centering
    \includegraphics[width=0.495\textwidth]{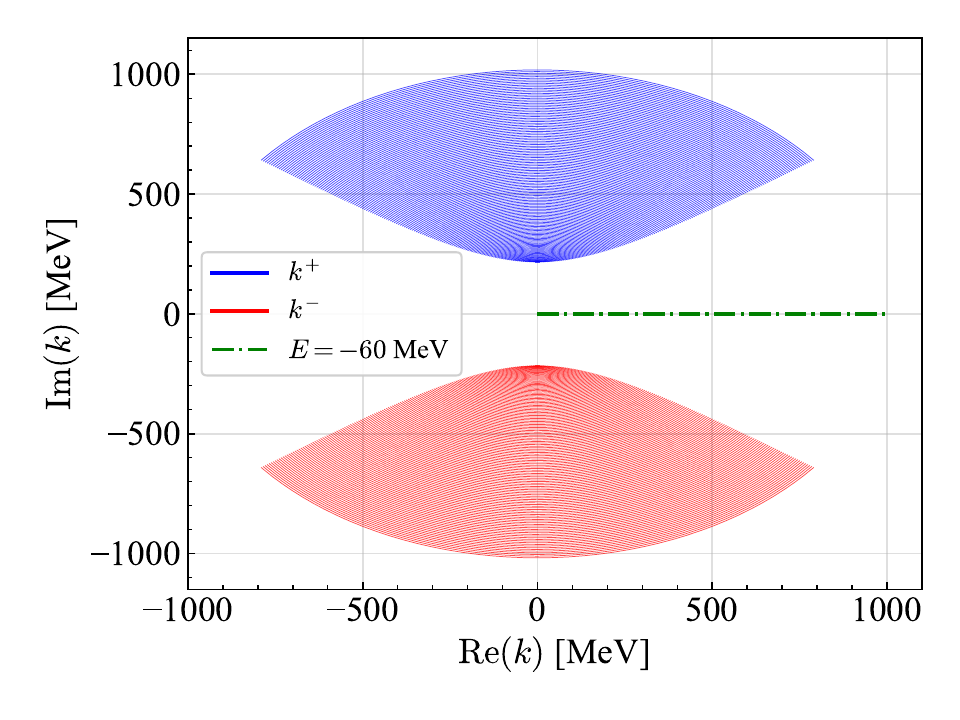}
    \includegraphics[width=0.495\textwidth]{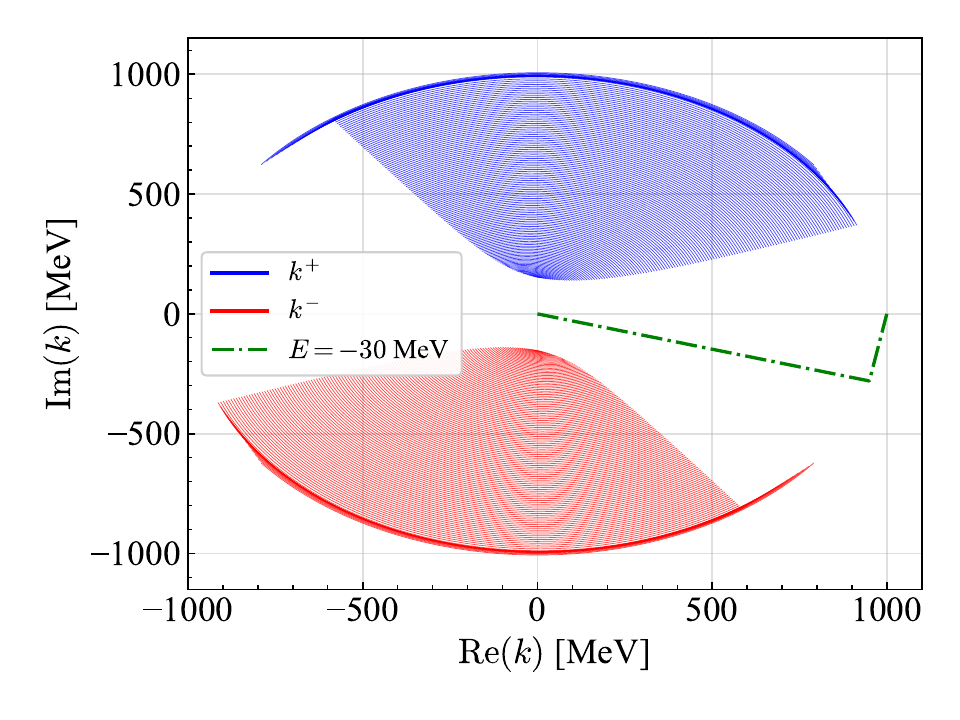}
    \caption{Plot of integration contours when $E$ is above $D_{s0}^{*}\bar{D}_{s}$ threshold ($-30$ MeV) and below threshold ($-60$ MeV) with the singularity caused by points taken from the contours. Green dashed dot lines are integration contours used in our calculation. The blue and red regions are composed of $k^{+}$ and $k^{-}$ of $V_{12}(p_{j},k)$ with $p_{j}$ moving along the integration contour.}
    \label{fig:K}
\end{figure}

Both of these singularities in the propagators and potentials will cause numerical instabilities as the integration is
performed along the positive real momentum axis. To overcome that problem, we use the contour deformation method as discussed in Refs.~\cite{Aaron:1966zz,Hetherington:1965zza,Cahill:1971ddy,BookThree}. First, the Faddeev equation is analytically continued in the complex momentum plane using Cauchy's theorem,
\begin{equation}
T(p,q)=V(p,q)+\int_{\mathcal{C}}K(p,k)T(k,q),
\label{eq:LSE2}
\end{equation}
where $\mathcal{C}$ denotes the deformed integration contour, and $p$ and $k$ are complex momenta. The $T$-matrix for the on-shell $p$ can then be obtained by using Cauchy's theorem for the second time. In this step, one must treat the singularities generated by the points on the integration contour, which form continuous regions built from the cut associated with each point on $\mathcal{C}$. In Fig.~\ref{fig:K}, choosing the integration along the contour $\mathcal{C}$ indicated by the green lines, the singularities in the $k$-plane when $E>-B_{DK}$ and $E<-B_{DK}$ are shown by the shaded areas. We can see that the Faddeev equation in Eq.~\eqref{eq:LSE2} is free of singularities for $E$ on the real axis, and can be solved by matrix inversion without crossing any singularity. This equation allows us to analytically continue $T(p,q_{\rm on})$ from $p\in\mathcal{C}$ to $T(q_{\rm on},q_{\rm on})$ safely when $E>-B_{DK}$; more details can be found in~\cite{Glockle:1978zz,Dawid:2023jrj}.

\end{document}